\documentclass{iau}
\usepackage{graphicx}

\title[JD 11.~~Statistical Challenges in 21st Century Cosmology] 
{Transformed Auto-correlation}

\author[Jianfeng Zhou \& Yang Gao]   
{Jianfeng Zhou$^{1, 2}$
 \and Yang Gao$^{1,2}$}

\affiliation{$^1$ Key Laboratory of Particle \& Radiation Imaging (Tsinghua University) \\[\affilskip]
$^2$Department of Engineering Physics and Center for Astrophysics,
Tsinghua University, Beijing 100084, China  
\\email: {\tt zhoujf@tsinghua.edu.cn, gaoyang12@mails.tsinghua.edu.cn}}

\pubyear{2014}
\volume{306}  
\pagerange{119--126}
\jname{Statistical Challenges in 21st Century Cosmology}
\editors{A.C. Editor, B.D. Editor \& C.E. Editor, eds.}
\begin{document}

\maketitle

\begin{abstract}
A transformed auto-correlation method is presented here,  where a received signal is transformed based on a priori reflecting  model, and then the transformed signal is cross-correlated to its original one. If the model 
is correct, after transformation, the reflected signal will be coherent to the transmitted signal, with zero delay. 
A map of transformed auto-correlation function with zero delay can be generated  in  a given parametric space.   
The significant peaks in the map may indicate the possible reflectors nearby the central transmitter. The true values of the parameters of reflectors  can be estimated at the same time.
\end{abstract}


\firstsection 
\section{Introduction}
Echo signals probably exist in the light curves of some sorts of astronomical objects, such as supernovae 
(\cite[Rest \etal\  2005]{rest2005}, \cite[Krause \etal\ 2008a]{krause2008a}, \cite[Krause \etal\ 2008b]{krause2008b}), x-ray binaries(\cite[Greiner 2001]{greiner2001}) and 
AGNs(\cite[Perterson 1993]{peterson1993}) etc.. If the relative location between a central transmitter and 
a reflecting object is fixed, then auto-correlation (\cite[Edelson \& Krolik 1988]{edelson1988}) can be used to detect the reflected signal and estimate the 
distance thereafter. However, in common situation, the nearby reflecting object is always moving 
(\cite[Perterson 2008]{peterson2008}), therefore, 
the delay between transmitted and reflected signal is variable, which means the coherence is broken and 
no echo signals could be detected by auto-correlation. 

Here, we are going to introduce a method which is called transformed auto-correlation. Its aim is to rebuild the coherence and perform cross-correlation between the reflected and transmitted signals.

\section{Basic Conception} 
Suppose there are only one transmitter and one reflector, and $S(t)$ 
is  a transmitted signal, and $R(t)$ is  the  received signal. Since astronomical objects are 
usually far far away, the received light curve is the combination of transmitted and reflected signals, i.e.,
\begin{equation}
R(t) = S(t) + \kappa S(t - \tau(t, a, b, c, ...))  \label{eqn1}
\end{equation}
where $\kappa$ is reflectance, and $\tau(t, a, b, c, ...)$                         
is the variable delay which is model dependent, $a, b, c, …$ are the parameters of the model.

Due to variable delay, a reflected signal will probably loss the coherence to the transmitted signal. 
In such situation, it is impossible to detect a reflected signal by auto-correlation. 
A transformation is needed to rebuild the coherence. Let's set :   
\begin{eqnarray}
\tilde{t} &=& t - \tau(t, a, b, c, ...)  \label{eqn2}\\
t &=& \Gamma(\tilde{t}, a, b, c, ...) \label{eqn3}
\end{eqnarray}
where $\Gamma()$ is the inversion function of $t$ in term of $\tilde{t}$. 

Applying Equation. \ref{eqn2} and \ref{eqn3} into Equation \ref{eqn1}, then we obtain a transformed 
received signal :
\begin{equation}
R^{T}(\tilde{t}) = S(\Gamma(\tilde{t}, a, b, c, ...)) + \kappa S(\tilde{t})
\end{equation}

{\bf The definition of transformed auto-correlation is the cross-correlation between 
received signal $R(t)$   and  its transformation  $R^{T}(t)$, i.e.}, 
\begin{equation} \label{eqn5}
C_{ta}(\tau, a, b, c, ...) = \big<R(t), R^{T}(t+\tau)\big> = \int R(t)R^{T}(t+\tau)dt 
\end{equation}

If the parameters are correct, then after transformation, the reflected signal is coherent to 
transmitted signal now, with zero delay. So, only $C_{ta}(0, a, b, c, ...)$  is important.
We can calculate all $C_{ta}(0, a, b, c, ...)$ in a given parametric space.  
The significant peaks of $C_{ta}(0, a, b, c, ...)$ may indicate possible reflectors existed nearby 
the central transmitter, and the true 
parameters of these reflectors cab be estimated at the same time. 

\section{Simulations}
Here, an one-dimension simulation is used as an example, where a transmitter is fixed at $x=0$, and several
reflectors are moving or fixed in x axis. The transmitted signal, generated by convolving a Gaussian white 
noise with a Gaussian point spread function ($\sigma=10$), is stochastic and band-limited.  The sampling interval is $1ms$.  and 2, 048, 000 sampled data are generated.  

Four reflectors are set in the simulation. The model parameters including Initial Position, Velocity and Reflectance 
are listed in Table \ref{tab1}, columns 2-4.

\begin{table}[!h]
\caption[]{The parameters of the Reflectors. Columns 2-4 are values of a model. Columns 5-7 are 
estimated values by transformed auto-correlation.} \label{tab1}
\begin{center}
\begin{tabular}{c|c|c|c||c|c|c}
\hline 

Reflector  &  Position$^1$(M)  & Velocity(M)  &  R$^2$(M) & \bf{Position(E)} & \bf{Velocity(E)} & \bf{R(E)}\\
                & (lc$^3$)  &  (c$^4$)  &                                                  & (lc)   & (c) & \\
\hline
1              &  0.8       &   -0.02   &  0.04   & \bf{0.80}   & \bf{-0.02}  & \bf{0.040}\\
2              &  0.9      &   0.0      &  0.08    & \bf{0.90}   & \bf{0.0}   & \bf{0.077 }\\
3              &  1.0      &   0.01    &  0.08    & \bf{1.00}  & \bf{0.01}  & \bf{0.078 }\\
4              &   1.2     &   0.03    &  0.06  & \bf{1.2} & \bf{0.03} & \bf{0.054 }\\
\hline
\end{tabular}
\end{center}
\vspace{1mm}
 \scriptsize{
 {\it Notes:}\\
  $^1$ Initial Position of the reflector.  $^2$ Reflectance. $^3$ Light Second. $^4$ Speed of Light.}
\end{table}

\begin{figure}[h]
\begin{center}
 \includegraphics[width=5in]{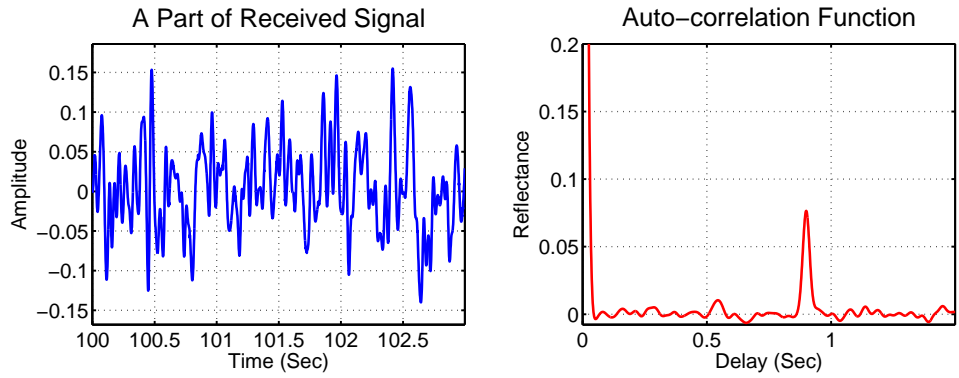} 
 \caption{A part of received signal(left) and its auto-correlation function(right).}
   \label{fig1}
\end{center}
\end{figure}

A part of received signal and its auto-correlation function are displayed in Figure \ref{fig1}.  
As shown in the auto-correlation function, although there are four 
reflectors, only one reflector, which is fixed at position of 0.9 light second, has a fringe. Other reflectors have 
variable delays, so the reflected signals are incoherent to the transmitted signals, and no fringes exist.

\begin{figure}[h]
\begin{center}
 \includegraphics[width=4in]{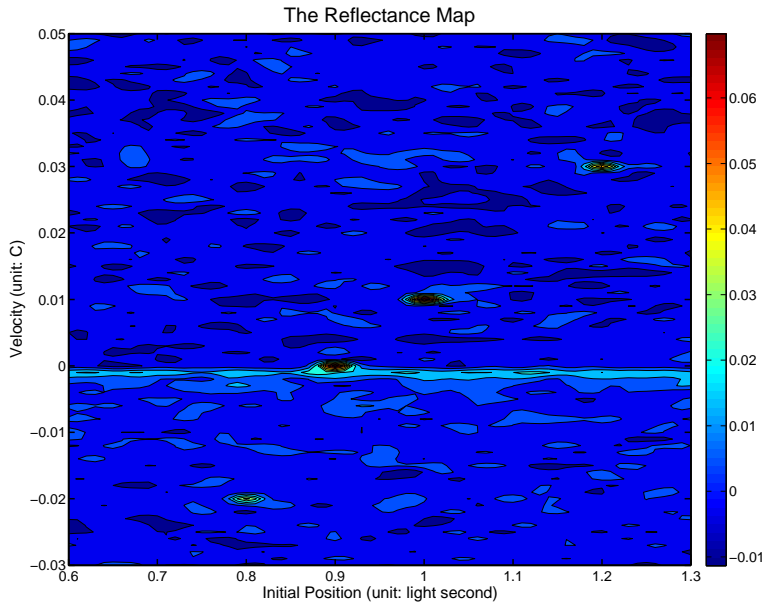} 
 \caption{Four reflectors with their initial positions and velocities are found 
by transformed auto-correlation method.}
   \label{fig2}
\end{center}
\end{figure}

Now, it is able to search the possible reflectors and their parameters by transformed auto-correlation. Here, 
$C_{ta}(0)$ is normalized by $C_{a}(0) = \big<R(t), R(t)\big>$. Thus the intensity of the peak of a reflector is 
equal to reflectance $\kappa$. The generated map in parametric space (initial position and velocity), which is called reflectance map,  is shown in Figure \ref{fig2}.   It is clearly 
seen that all of four reflectors have been detected. The estimated
parameters (initial position, velocity and reflectance) are listed in Table \ref{tab1}, columns 5-7.  The standard
deviation of the background in reflectance map is about 0.0036.

\section{Conclusions}
Transformed auto-correlation is able to detect reflected signals when there are constant or variable delays. With a priori reflecting model,  the parameters (such as position, velocity etc..) of reflectors can also be estimated.

\end{document}